\documentclass{ws-procs9x6}

\newcommand{\De}{\Delta}

\newcommand{\Om}{\Omega}

\newcommand{\p}{\partial}
 
\newcommand\eqn[1]{(\ref{#1})}      
\newcommand{\beq}{\begin{equation}}
\newcommand{\eeq}{\end{equation}}
\newcommand{\ba}{\begin{array}}
\newcommand{\ea}{\end{array}}

\newcommand{\MeV}{{\rm MeV}}

\begin{document}

\title{Gapless CFL and its competition with mixed phases\footnote{
\uppercase{T}alk given by \uppercase{C}. \uppercase{K}ouvaris.
}
}

\author{Mark Alford
}

\address{Physics Department, Washington University,
St.~Louis, MO~63130, USA
 }

\author{Chris Kouvaris and Krishna Rajagopal}

\address{Center for Theoretical Physics, Massachusetts 
Institute of Technology, Cambridge, MA 02139, USA
}  

\maketitle

\abstracts{ 
We recently argued that as the density of quark
matter decreases, there is a continuous transition from
the color-flavor-locked (CFL) phase to a gapless ``gCFL'' phase.
The reason is the growing importance of the strange quark mass,
and the constraint of electric/color neutrality.
In this paper we discuss mixed phases that achieve neutrality by charge
separation, and might offer an
alternative to the gCFL phase.
We find that none of the obvious mixtures
is favored relative to gCFL.
}

\section{Introduction}

There is a strong possibility that quark matter
may occur in the core of compact (``neutron'') stars, and at sufficiently
high density it is expected to be color superconducting:
the attractive QCD interaction between the quarks at the Fermi surface
causes them to condense in Cooper pairs.
At asymptotic densities, where the up, down and strange
quarks can be treated on an equal footing and the potentially disruptive
strange quark mass can be neglected, quark matter is in
the  color-flavor locked (CFL) phase~\cite{Alford:1998mk}, in which quarks of 
all three colors and all three 
flavors form Cooper pairs~\cite{reviews}.


However a very interesting question is what happens at lower densities.
In dense matter with quark chemical potential $\mu$ and pairing gap $\De$,
the strange quark mass has negligible impact on CFL pairing when
$M_s^2\ll \mu\De$. At compact-star densities $\mu\sim 350$ to $500~\MeV$,
$M_s\sim 80$ to $500~\MeV$, and
$\De\sim 10$ to $100~\MeV$, so it is quite possible that the strange quark
has a strong influence on the pairing.

Starting at asymptotic density and reducing the density
by reducing $\mu$, we expect CFL pairing to be disrupted,
at first by kaon condensation\cite{Bedaque:2001je} (which we do not
discuss), and then by the transition
to gapless CFL (gCFL)\cite{Alford:2003fq,Alford:2004hz}
with the appearance of gapless modes.
In this paper we assume that this happens before the
transition from quark matter to hadronic matter.
We first review some of the features of gCFL, 
and then compare it with
various charge-separated mixed phases, finding that
they are not energetically favored
even before we include their extra energy costs
arising from Coulomb (color-)electric fields and the
surface tension.

\section{The Gapless CFL Phase}

Stable bulk matter has to be neutral under all the gauged symmetries
(color and electromagnetic), and equilibrated under all interactions
including the weak interaction. These conditions are imposed by
introducing chemical potentials coupled to the gauged charges: 
$\mu_e$ couples to negative
electric charge, and $\mu_3$, $\mu_8$ couple to the color generators
$T_3$ and $T_8$, generators of the Cartan subalgebra of the color group.
$\mu_e, \mu_3$, $\mu_8$ are determined by
requiring that the corresponding charge densities vanish.
The pairing ansatz we use is
\begin{equation}
\langle \psi^\alpha_a C\gamma_5 \psi^\beta_b \rangle \sim 
\Delta_1 \epsilon^{\alpha\beta 1}\epsilon_{ab1} \!+\! 
\Delta_2 \epsilon^{\alpha\beta 2}\epsilon_{ab2} \!+\! 
\Delta_3 \epsilon^{\alpha\beta 3}\epsilon_{ab3}
\label{condensate}
\end{equation} 
The gap parameters
$\Delta_1$, $\Delta_2$ and $\Delta_3$ describe down-strange,
up-strange and up-down Cooper pairs, respectively.
Above a critical $M_s^2/\mu=2\Delta$, the CFL phase 
is replaced by a new gapless CFL (gCFL) phase\cite{Alford:2003fq}.
The defining properties of the
gapless CFL phase arise in its dispersion relations,
not in its pattern of gap parameters.  However, it is
useful for orientation to list the patterns of 
gap parameters for all the phases we shall discuss:
\begin{equation}
\ba{r@{\,\,}l@{\qquad}r@{\,\,}l}
 \mbox{CFL}:& \Delta_3 \!\simeq\! \Delta_2 \!=\! \Delta_1 \!=\!\Delta_{CFL}
& \mbox{gCFL}: & \Delta_3 > \Delta_2 >\Delta_1>0 \\
\mbox{g2SC}: & \Delta_3>0,\quad  \Delta_1 \!=\! \Delta_2 \!=\! 0
& \mbox{2SCus}: & \Delta_2>0,\quad  \Delta_1 \!=\! \Delta_3 \!=\! 0
\ea
\label{phases}
\end{equation}
The 2SCus phase has the same free energy as 2SC
at $M_s=0$, and to leading order in
$M_s$ if their respective nonzero
gap parameters have the same value\cite{Alford:2002kj}.
However, an NJL model calculation 
indicates that $\De_2$ in 2SCus is always less than $\De_3$ in 2SC,
so 2SCus is never favored\cite{Alford:2004hz}.

\section{Mixed phases}

\begin{figure}[ht]
\centerline{\epsfxsize=4.1in\epsfbox{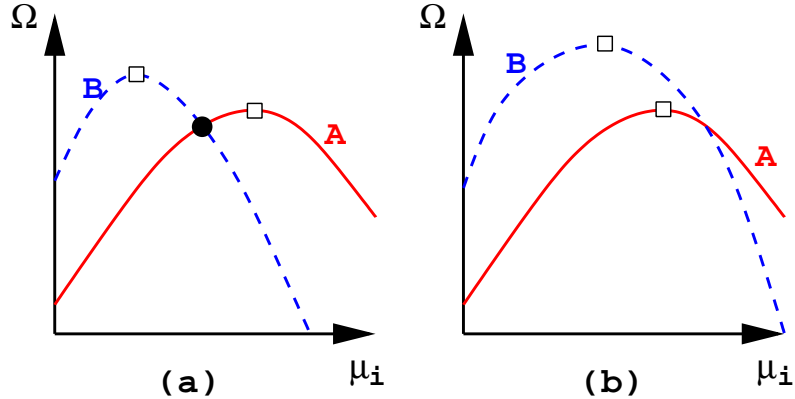}}   
\caption{Schematic illustration of the dependence of the free energy $\Om$
on a gauged chemical potential $\mu_i$, showing
conditions for the occurrence of mixed phases. 
Charge $Q_i=-\partial\Omega/\partial\mu_i$
is given by the slope. Squares mark the neutral points.
Panel~(a): at the neutral value of $\mu_i$ for each phase,
the other phase has lower free energy,
so there is a coexistence point (black dot) for oppositely-charged phases
with lower free energy than either neutral phase.
Depending on Coulomb and
surface energy costs, a mixed phase may exist there.
Panel~(b): phase $B$ has higher free energy
than phase $A$ at the value of $\mu_i$ where $A$ is neutral. 
The two phases never coexist with opposite charge, so no mixed phase
is possible.}
\label{fig:mixed}
\end{figure}

The gCFL phase can be viewed as a distortion of the CFL phase, induced
by the stress of a non-zero strange quark mass, combined with the
constraints of color and electromagnetic neutrality. However, it is possible
for neutrality to be achieved by a mixture of oppositely charged phases
at the same pressure. This happens in
the two-flavor case, where a gapless 2SC phase exists\cite{Shovkovy:2003uu},
but a mixed phase is free-energetically favored\cite{Reddy:2004my}.
We now ask whether such a mixed
phase might be preferred over the gCFL phase.

The two possible situations are schematically illustrated in 
Fig.~\ref{fig:mixed},
which shows generic free energy curves $\Omega(\mu_i)$ for two phases 
$A$ and $B$. In Fig.~\ref{fig:mixed}b
there is no  coexistence point and hence no mixed phase is possible.
In Fig.~\ref{fig:mixed}a 
there is a coexistence point of oppositely-charged
phases, and its free energy is lower than that 
of either neutral phase, so if the energy of the electric fields induced by
the charge separation and the surface tension of the interface is small enough,
a neutral mixed phase will be free-energetically preferred
over either homogeneous neutral phase.

In quark matter we have gauge charges associated with color as well as
electromagnetism. However, it seems unlikely that mixed phases involving
color-charged components could form. The color gauge coupling 
is strong, so separating color charges into different domains will produce
color-electric fields with very high energy cost.
It seems likely, therefore, that mixed phases will consist of components
that are individually color-neutral, with different values of the color
chemical potentials $\mu_3$ and $\mu_8$, but which are electrically
charged, with a common electric chemical potential $\mu_e$.
In this case there are still color-electric energy costs, associated with the
color electric field induced at the phase boundary by the gradient
in $\mu_3$ and $\mu_8$. We will not try to calculate these interface costs
\cite{Alford:2001zr,Reddy:2004my}, since our argument is that
most of the mixed phases 
are excluded even before we include the electrostatic and
interface energy costs.

In the region $(M_s^2/\mu)>2\De$
where gCFL is the free-energetically favored homogeneous neutral
phase, the other less-favored quark matter phases are
2SC, 2SCus, and unpaired quark matter \eqn{phases}.
Note that there is no CFL solution, charged or neutral, in this region.
We have compared the mixed phases with neutral gCFL. 
We performed calculations in an NJL
model at $\mu=500~\MeV$, with coupling chosen so that $\De_{CFL}=25~\MeV$
at $M_s=0$\cite{Alford:2003fq,Alford:2004hz}. 
When we vary $M_s$ at fixed $\mu$ we find the gCFL region is
$47~\MeV<M_s^2/\mu<130~\MeV$. At the upper limit there is a convergence of the
free energies of g2SC, gCFL, and unpaired quark matter, so 
there is a first-order transition from gCFL to unpaired, with a possible
window of g2SC\cite{Ruster:2004eg}.
We have performed calculations of the various
possible mixtures. 
We now discuss the results.

\newcounter{item}
\setcounter{item}{0}
\addtocounter{item}{1}
\noindent(\theitem) \underline{unpaired+gCFL}.
For color neutral unpaired and gCFL phases,
the situation is typically that of Fig.~\ref{fig:mixed}(b)
so mixed phases are ruled out.
This is true for all values of $M_s^2/\mu$
except for a range of a few MeV just below $M_s^2/\mu=130$~MeV,
where the neutral gCFL and neutral unpaired free energies cross.
There, a mixed phase may arise, although 
it may be superseded by other more favorable possibilities
such as the crystalline 
phase\cite{LOFF}.

\addtocounter{item}{1}
\noindent(\theitem) \underline{2SC+2SCus}.
We find the situation of Fig \ref{fig:mixed}a: a neutral mixed phase
exists. However, its free energy is higher than that of gCFL
even before electrostatic and surface energy costs are included.
At $M_s^2/\mu=80~\MeV$, it has
$\Om=-14.93\times 10^6~\MeV^4$, vs.~$\Om_{\rm gCFL}=-18.01\times 10^6~\MeV^4$. 
(These free energies are both measured relative to that of neutral 
unpaired quark matter.)
We have checked that a neutral 2SC+2SCus mixed phase is
also free-energetically unfavored relative to homogeneous gCFL
at $M_s^2/\mu=51.2~\MeV$,
which is just above the CFL$\to$gCFL transition.

\addtocounter{item}{1}
\noindent(\theitem) \underline{gCFL+2SCus}.
The $\mu_e$ dependence of 
the free energies of these two phases is as in Fig \ref{fig:mixed}b,
so a mixed phase is not possible.
We have verified this at $M_s^2/\mu=51.2~\MeV$ and $80~\MeV$.

\addtocounter{item}{1}
\noindent(\theitem) \underline{gCFL+2SC}.
At $M_s^2/\mu=51.2~\MeV$ we find the situation of Fig.~\ref{fig:mixed}b,
so no mixed phase is possible. At $M_s^2/\mu=80~\MeV$,
the free-energy dependence is of the type shown in Fig.~\ref{fig:mixed}a,
so a mixed phase is possible, and since gCFL itself is one of the components
its free energy is necessarily lower than that of neutral gCFL.
However, $\Om_{gCFL}(\mu_e)$ depends {\em very} weakly on $\mu_e$: the
$\Om_{gCFL}(\mu)$ parabola is very shallow. This suppresses the
mixed phase in two ways.
(i) the free energy of the mixed phase is only lower than that of
neutral gCFL by a very small margin ($0.0012\times 10^{6}~\MeV^4$ at
$M_s^2/\mu=80~\MeV$), so there is very little chance that it will
survive once electrostatic and surface costs are included;
(ii) The charge density of
gCFL, proportional to $\p\Om/\p\mu_e$, is very small
so the mixed phase must be
dominantly gCFL, with a tiny admixture of 2SC, to achieve neutrality.
It is known that such highly asymmetric mixed phases have the
highest electrostatic energy costs\cite{Alford:2001zr}.

We conclude that the gCFL phase is a very strong candidate for the ``next 
phase down in density'' after CFL. None of the mixed phases that we 
investigated was able to compete with it. It remains to be seen what
role kaon condensation (in CFL and gCFL)
plays, and whether a completely different phase
such as the crystalline (LOFF) phase\cite{LOFF} can compete with gCFL 
at sufficiently low density.

{\bf Acknowledgements}: We thank J. Bowers for helpful conversations.
This research was supported in part by DOE grants 
DE-FG02-91ER40628 and DF-FC02-94ER40818.

\end{document}